\documentclass[twocolumn,letter]{jpsj2} %% two-column layout
\usepackage{slashbox}

\title{
Fermi-Suface Evolution by Transition-metal Substitution in the Iron-based
Superconductor LaFeAsO 
}

\author{Shun \textsc{Konbu}$^{1}$, Kazuma \textsc{Nakamura}$^{1,2}$, Hiroaki \textsc{Ikeda}$^{4}$, Ryotaro \textsc{Arita}$^{1,2,3}$}
\inst{$^1$Department of Applied Physics, University of Tokyo, 7-3-1 Hongo, Bunkyo-ku, Tokyo, 113-8656, Japan \\ 
$^2$JST TRIP/CREST, 7-3-1 Hongo, Bunkyo-ku, Tokyo, 113-8656, Japan \\ 
$^3$JST-PRESTO, Kawaguchi, Saitama 332-0012, Japan \\ 
$^4$Department of Physics, Kyoto University, and JST TRIP, Sakyo-ku, Kyoto, 606-8502, Japan} 

\recdate{\today}

\abst{
We study how Co- and Ni-substitution affect the electronic structure of the iron-based superconductor, LaFeAsO. We perform {\it ab initio} supercell calculations and unfold the first Brillouin zone (BZ) to calculate the spectral function in the BZ for the normal cell. The charge density distribution in real space shows that doped extra electrons are trapped around Co (Ni) atom. This seems to suggest that carriers are not doped by Co(Ni)-substitution. However, the present momentum-space analysis indicates that the Fermi-surface volume indeed expands by substitutions, which can be well described by the rigid-band shift approximation. By taking into account this effective doping, we discuss whether the sign-reversing $s$-wave ($s_{\pm}$-wave) scenario is compatible with experiments.
}

\kword{iron-based superconductor, impurity effect, pairing mechanism}

\begin{document}
\maketitle 

Since the discovery of superconductivity in F-doped LaFeAsO~\cite{Kamihara}, iron-based superconductors are now attracting broad interests due to their high-transition temperature ($T_c$) up to 56 K~\cite{iron-review}. It has been experimentally found that we can enhance or even induce the high $T_c$ superconductivity by several ways such as chemical substitution, (intentional) introduction of defects, application of external pressure, and so on. Among them, transition-metal substitution is of great interest and has been intensively studied both theoretically and experimentally, since it has two important effects from which we may get crucial hints to clarify the pairing mechanism: One is carrier doping effect and the other is pair breaking effect.

Theoretically, it has been shown that the local potential of transition metal ions has a systematic chemical trend \cite{Nakamura}, and the position of the Fermi level ($E_F$) in the density of state (DOS) exhibits a monotonic shift as a function of the concentration of extra electrons donated by the transition-metal substitutes~\cite{Nakamura,Singh}. On the other hand, the charge density distribution $\rho ({\bf r})$ of extra electrons are localized around the impurity site~\cite{Wadati}, which seemingly suggests that transition-metal substitution does not dope mobile carriers into the system. While these two observations seem to contradict with each other, it should be noted that the structure of $\rho ({\bf r})$ does not necessarily give conclusive information on whether the extra electrons are localized or not. Due to the local charge neutrality condition, $\rho({\bf r})$ always has a peak structure around the impurity site, whose amplitude is determined by the Friedel sum rule. Experimentally, systematic analysis of transport properties suggests that carriers are indeed doped into the system.~\cite{Canfield,Sato} Co-, Rh-, Ni- and Pd-doping dependence of $T_c$ were also investigated systematically, and it was found that $T_c$ seems to depend only on the amount of extra electrons. This behavior can be understood if the charge-doping effect is appreciable, but the pair-breaking effect is negligibly small.

A standard way to study impurity effects by first-principles calculation is to introduce a large supercell and replace one of the atoms in the cell with a dopant atom. Although this approach can mimic a randomly doped system when the size of supercell is sufficiently large~\cite{DOS}, it also has a drawback; the first Brillouin zone (BZ) for the normal cell are folded into the small BZ, where many bands are heavily entangled. Regarding this problem, recently, a technique of BZ-unfolding was developed~\cite{Ku}, which enable us to transform the band structure in the original small BZ for the supercell to that in the BZ for the normal cell. 

In the present study, using this method, we examine the effect of Co- and Ni-doping on the electronic structure. By visualizing the evolution of the Fermi-surface volume by transition-metal substitution, we show that the system is effectively doped, although the charge density of extra electrons are localized around the transition metal impurity ions. We estimate the amount of doped carriers and show that this doping is well described by a rigid-band shift of the Fermi level. Finally, we discuss the impurity effect on the superconductivity.

Let us move on to the detail of the method employed in the present study. First, we performed {\it ab initio} density-functional calculations with generalized-gradient-approximation (GGA) \cite{PBE}, using {\it Tokyo Ab initio Program Package} \cite{TAPP} with a plane wave basis set and norm-conserving pseudopotentials \cite{Ref_PP}. The cutoff energy for the wave functions and charge density were set to be 64 Ry and 256 Ry, respectively. We replaced one Fe atom in 2$\times$1$\times$1, 2$\times$2$\times$1 and 3$\times$3$\times$1 supercell with Co or Ni \cite{DOS}. Since the normal cell contains two Fe atoms, these setups correspond to 25 \%, 12.5 \% and 5.6 \% doping, respectively. While a $k$-mesh of 5$\times$5$\times$5 was used for the former two supercells, a 3$\times$3$\times$3 $k$-mesh was used for the latter supercell. 

Next, we constructed maximally localized Wannier functions (MLWFs) \cite{MaxLoc} for the Fe 3$d$ bands around the energy range of $-$3 eV to 3 eV. We then represent the Kohn-Sham Hamiltonian ${\cal H}^{\rm KS}$ using the resulting supercell Wannier orbitals as ${\cal H}^{\rm KS}_{NN'}({\bf R},{\bf R'})=\langle N {\bf R}|{\cal H}^{\rm KS}|N' {\bf R'} \rangle$, where $|N {\bf R} \rangle$ denotes the $N$-th Wannier orbital sitting at the lattice ${\bf R}$. Note that ${\cal H}^{\rm KS}_{NN'}({\bf R},{\bf R'}) = {\cal H}^{\rm KS}_{NN'}({\bf R-R'})$ due to the translational symmetry. The advantage of the MLWF method is that we can directly estimate microscopic quantities in real space, associated with the impurity effect \cite{Nakamura}. 

Table~\ref{I_nimp_rho} summarizes estimated onsite impurity potential $I$ (onsite-level difference between impurity and Fe atoms) and occupancy at the impurity site $n_{\rm imp}$ for the three impurity concentrations. We found that the $I$ values have weak concentration dependence and were estimated to be $\sim$$-$0.4 eV for Co and $\sim$$-$0.9 eV for Ni. The occupancy $n_{\rm imp}$ is nearly 7.0 for Co, and 7.7 for Ni, apparently suggesting that extra $d$ electrons are trapped at the impurity site. This is consistent with the result in Ref.~\cite{Wadati}. However, real-space analysis is not convenient to see the carrier doping effect. By visualizing the evolution of the Fermi-surface volume, below we will estimate the amount of carriers doped into the system effectively.

\begin{table}[htb] 
\caption{Estimated onsite impurity potential $I$ (eV) and occupancy of the impurity site $n_{{\rm imp}}$ for LaFe$_{1-x}$M$_{x}$AsO.} 

\
 
\centering 
%{\scriptsize %%%%%%%%%%%%%%%%%%%%%%%%
\begin{tabular}{
c@{\ \ \ }r@{\ \ \ }
r@{\ \ \ }r@{\ \ \ }r@{\ \ }
r@{\ \ \ }r@{\ \ \ }r@{\ \ }} 
 \hline \hline \\ [-5pt]
       & \multicolumn{3}{c}{M = Co} &        
       & \multicolumn{3}{c}{M = Ni} \\ [2pt] \hline \\ [-8pt] 
$x$ =  & 0.056 & 0.125 & 0.250 &        
       & 0.056 & 0.125 & 0.250 \\ 
[2pt] \hline \\ [-8pt]
%$\bar{\Omega}$  & 2.03 & 1.97 & 2.00 &    & 2.39 & 2.29 & 2.30 \\
$I$   & $-$0.35 & $-$0.37 & $-$0.44 &  
      & $-$0.86 & $-$0.88 & $-$1.05 \\ 
$n_{{\rm imp}}$ & 6.97 & 6.97 & 6.99 & 
                & 7.70 & 7.72 & 7.81 \\
%$\rho(E_F)$ & 12.50 & 8.76 & 7.41 &
%            &  7.96 & 7.97 & 9.92 \\ 
\hline \hline \\ 
\end{tabular} 
%} 
\label{I_nimp_rho} 
\end{table}

Recently, Ku {\it et al.} presented a scheme to unfold bands in the first BZ for the supercell by utilizing the Wannier functions \cite{Ku}. The spectral function 
%does not depend on the choice of basis of the one-particle Green function, and it 
is calculated as 
\begin{eqnarray}
A({\bf k},\omega) = -\frac{1}{\pi} \sum_{n} {\rm Im} G_{nn}({\bf k},\omega),
\label{Akomega}
\end{eqnarray}
where ${\bf k}$ is wavevector in the unfolded BZ, $\omega$ is frequency, and $G_{nn}({\bf k},\omega)$ is the Green function in the Wannier basis $n$ in the primitive cell. Unless the impurity changes the electronic structure drastically, the quantity ${\rm Im} G_{nn}({\bf k},\omega)$ can be obtained from the supercell calculation including the impurity as  
\begin{eqnarray}
{\rm Im} G_{nn} ({\bf k},\omega) = \sum_{{\bf K}J}|\langle n{\bf k} | J{\bf K} \rangle|^2 \delta(\omega - E_{J{\bf K}}),
\label{ImGnn}
\end{eqnarray} 
where $|J{\bf K} \rangle$ is the $J$-th Bloch state with wavevector ${\bf K}$ in the supercell BZ and $E_{J{\bf K}}$ is the corresponding eigenvalue. The expression for $\langle n{\bf k} | J{\bf K} \rangle$ is found in Ref.~\cite{Ku}. 

\begin{figure*}[htbp]
\vspace{0cm}
\begin{center}
\includegraphics[width=0.9\textwidth]{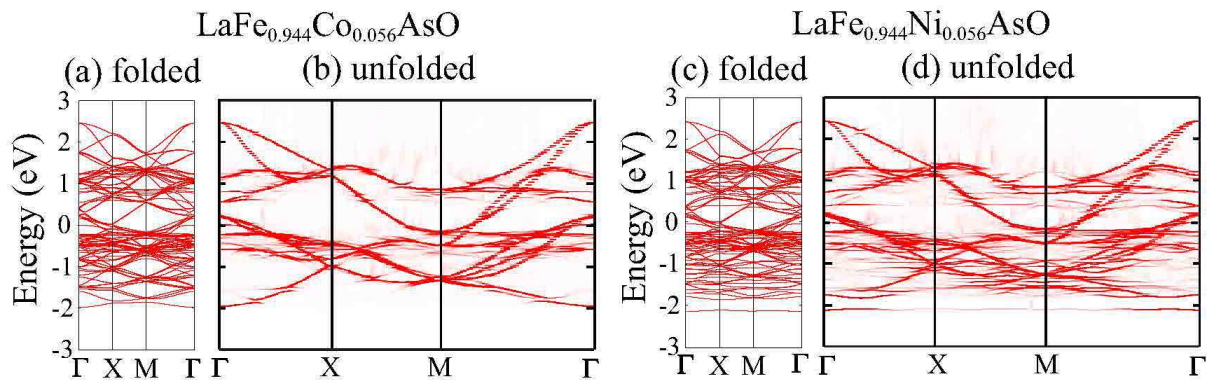}
\caption{(Color online)
(a) Band structure of the effective model for the $d$ bands of LaFe$_{0.944}$Co$_{0.056}$AsO in a 3$\times$3$\times$1 supercell. The band dispersion reproduces that of GGA. Due to band-folding, there are 90 bands in the first BZ. (b) Unfolded band structure. Due to the impurity effect, there are many gaps and several shadow bands in the dispersion. (c) A plot similar to panel (a) for LaFe$_{0.944}$Ni$_{0.056}$AsO. (d) Unfolded band structure. The size of gaps and the number of shadow bands are larger than those in the Co case.
}
\label{Fig1}
\end{center}
\end{figure*}

Now, let us look into the results for impurity doping. Figures~\ref{Fig1}~(a) and (b) [(c) and (d)] describe the Co (Ni) case.  In the panels (a) and (c), the band dispersions for the 3$\times$3$\times$1 supercell (corresponding to 5.6 \% doping) are plotted. The total number of transition-metal atoms in the cell is 18, so that we have 90 $d$ bands. 
The panels (b) and (d) show spectral functions $A({\bf k},\omega)$ [Eq.~(\ref{Akomega})] for the normal cell obtained after the BZ unfolding. The effect of impurity doping appears as small gaps or shadow bands in the band dispersion. As we will see later, the position of $E_F$ also shifts. 
For the same doping concentration (5.6\%), the effect of Ni-doping is more significant than that of Co-doping. However, for both the systems, the band structure does not change drastically. This is because the size of the impurity potential ($<$ 1 eV) is still small compared to the bandwidth of the $d$ bands ($\sim$ 4.5 eV). 

\begin{figure}[htbp]
\vspace{0cm}
\begin{center}
\includegraphics[width=0.4\textwidth]{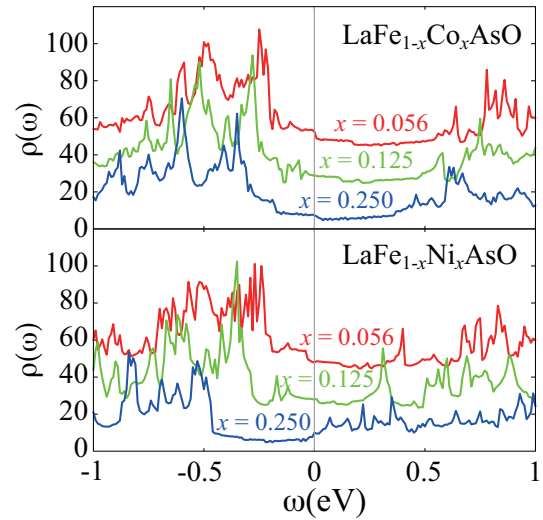}
\caption{(Color online) $x$ dependence of the DOS $\rho(\omega)$ in units of (States/eV): (Upper panel) LaFe$_{1-x}$Co$_{x}$AsO; (lower panel) LaFe$_{1-x}$Ni$_{x}$AsO. The base lines for $x=0.056$ and $0.125$ are shifted by 40 and 20, respectively.} 
\label{Fig2}
\end{center}
\end{figure}

We show in Fig.~\ref{Fig2} electronic density of states $\rho(\omega) = \sum_{{\bf K}J} \delta (\omega - E_{J{\bf K}})$ for LaFe$_{1-x}$Co$_x$AsO (upper panel) and LaFe$_{1-x}$Ni$_x$AsO (lower panel) with various impurity concentrations \cite{DOS}. From the figure, we see that $\rho(\omega)$ around $E_F$ ($-$0.5 eV to 0.5 eV) shows a monotonic downward shift with increasing impurity concentration. This behavior can be understood in terms of the center of mass of $\rho(\omega)$; since the onsite potential of Co and Ni are deeper than that of Fe, the center of $\rho(\omega)$ for the impurity system shifts downward compared to the undoped system. As the impurity concentration increases, this shift becomes larger. We note that the concentration dependence of $\rho(\omega)$ is more significant for Ni than Co, due to the fact that the former onsite potential is deeper than the latter one (see Table~\ref{I_nimp_rho}). Indeed, $\rho(\omega)$ of the Ni-doped system is strongly distorted from that of the pure system.

\begin{figure*}[htbp]
\vspace{0cm}
\begin{center}
\includegraphics[width=0.8\textwidth]{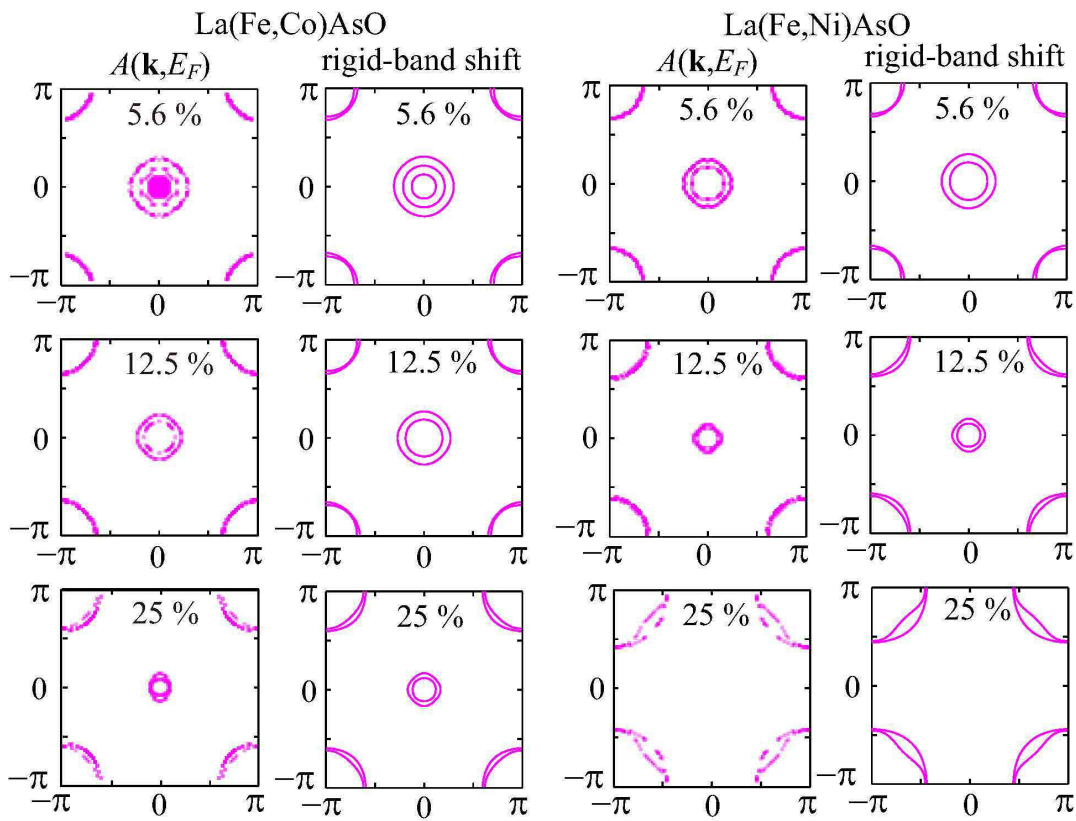}
\caption{(Color online) Fermi surface evolution by impurity-doping obtained by the BZ unfolding method (Co: left two panels; Ni: right two panels). The figures compare $A({\bf k},E_F)$ in Eq.~(\ref{Akomega}) with the Fermi-surface obtained by the rigid-band shift approximation. Doping concentration is 5.6, 12.5 and 25 \% from top to bottom.} 
\label{Fig3}
\end{center}
\end{figure*}

In Fig.~\ref{Fig3}, we show how the Fermi surface changes by Co doping (left two panels) and Ni one (right two panels). For 5.6, 12.5 and 25 \% concentrations, we compare $A({\bf k},\omega=E_F)$ obtained by the BZ-unfolding method (left side in the two panels) with the Fermi surface obtained by the shift of the Fermi level of the pure system (right side). We see that the size and shape of Fermi surfaces are quite similar to each other.\cite{C4} This clearly indicates that substitution of Co/Ni impurities dope carriers into the system effectively. We can expect that such Fermi-surface evolution exists not only in La1111 system, but also in other systems such as Ba122.

Finally, let us discuss the impurity effect on superconductivity. Among various pairing symmetries, the possibility of pairing with the $a_{1g}$ symmetry has been extensively studied. When the onsite pairing mediated by orbital/charge fluctuations is dominant, the gap function does not have sign changes in momentum space (the so-called $s_{++}$-pairing)~\cite{Kontani2}. On the other hand, the gap function has sign changes (the so-called $s_{\pm}$-wave) when the off-site pairing mediated by spin fluctuations is dominant.~\cite{s+-} 

Experimentally~\cite{Canfield,Sato}, as mentioned above, $T_c$ seems to depend only on the number of extra electrons of the transition metal impurities. In the $s_{++}$-pairing scenario, due to the Anderson theorem, the pair breaking effect by non-magnetic impurities is negligibly small~\cite{Kontani}. Therefore, if we assume that the $s_{++}$-pairing interaction is suppressed by the Fermi surface evolution, we can naturally understand why the reduction of $T_c$ depends only on the number of extra electrons.

On the other hand, the Fermi-surface evolution can be fatal also for the $s_{\pm}$ pairing, especially when the Fermi surface deforms too much and spin fluctuations are suppressed. Therefore, to explain the experiments, the pair breaking effect for the $s_{\pm}$-pairing should be also weak. Regarding this problem, Ref.~\cite{Kontani} has shown that the pair breaking effect is appreciable for the $s_{\pm}$ pairing. However, the following points should be noted: First, the pair-breaking effect depends sensitively not only on the depth of the scattering potential, but also on its sign \cite{Kontani}. The superconductivity is very fragile against impurity doping for positive scattering potential, but rather robust for negative potential. As seen in Table~\ref{I_nimp_rho}, Co- and Ni-impurity potential are negative. We should also note that the competition between the on-site and off-site pairing depends sensitively on parameters/terms in the effective model and how we solve that model. In the random-phase approximation (RPA) for the Hubbard model, the contribution of the on-site component is estimated to be small \cite{Kariyado}. However, with treating the electron correlations beyond the RPA level or introducing electron-phonon couplings in the model \cite{Kontani}, we may have situations where the on-site contribution to the $s_{\pm}$-wave pairing becomes larger. It is an interesting future problem to examine these points quantitatively.

In summary, we have studied how Co- and Ni-substitution change the electronic band structure of the iron-based superconductor, LaFeAsO. We found that mobile carriers are indeed doped into the system, while the charge density distribution seemingly suggests that extra electrons are trapped around the impurity atoms. We demonstrated unambiguously such effective carrier doping by visualizing the Fermi-surface evolution. The doping effect is well described by a rigid-band shift of the Fermi level. Based on these results we discussed how the $s$-wave ($s_{\pm}$-wave) scenario can be compatible with experiments.

We thank I. Elfimov, M. Harverkort, P. Hirschfeld, I. Mazin and G. Sawatzky for fruitful discussion. Calculations were done at Supercomputer center at Institute for Solid State Physics, University of Tokyo. This work was supported by Grants-in-Aid for Scientific Research (No.~22740215, 22104010, and 23110708), Funding Program for World-Leading Innovative R\&D on Science and Technology (FIRST program) on ``Quantum Science on Strong Correlation'', JST-PRESTO and the Next Generation Super Computing Project and Nanoscience Program from MEXT, Japan.


\begin{thebibliography}{99}
\bibitem{Kamihara}Y. Kamihara, T. Watanabe, M. Hirano, and H. Hosono: J. Am. Chem. Soc {\bf 130} (2008) 3296.
\bibitem{iron-review} For a recent review on iron-based superconductors, see e.g., J. Paglione, and R. L. Greene: Nature Phys. {\bf 6} (2010) 645.
\bibitem{Nakamura}K. Nakamura, R. Arita, and H. Ikeda: Phys. Rev. B {\bf 83} (2011) 144512.
\bibitem{Singh} A. S. Sefat, R. Jin, M. A. McGuire, B. C. Sales, D. J. Singh, and D. Mandrus: Phys. Rev. Lett., {\bf 101} (2008) 117004.
\bibitem{Wadati}H. Wadati, I. Elfimov, and G.A. Sawatzky: Phys. Rev. Lett., {\bf 105} (2010) 157004.
\bibitem{Canfield}N. Ni, A. Thaler, A. Kracher, J. Q. Yan, S. L. Bud'fko, and P. C. Canfield: Phys. Rev. B {\bf 80} (2009) 024511.
\bibitem{Sato} T. Kawamata, E. Satomi, Y. Kobayashi, M. Itoh, and M. Sato: J. Phys. Soc. Jpn., {\bf 80} (2011) 084720. 
\bibitem{DOS}
The present calculations do not consider configuration averages over the impurity position, which will generate smearing of the density of states.
\bibitem{Ku}W. Ku, T. Berlijn, and C-C. Lee: Phys. Rev. Lett. {\bf 104} (2010) 216401. For more recent application, see T. Berlijn, D. Volja, and W. Ku: Phys. Rev. Lett. {\bf 106} (2011) 077005. 
\bibitem{PBE}J. P. Perdew, K. Burke, and M. Ernzerhof: Phys. Rev. Lett. {\bf 77} (1996) 3865.
\bibitem{TAPP}J. Yamauchi, M. Tsukada, S. Watanabe, and O. Sugino: Phys. Rev. B {\bf 54}  (1996) 5586. 
\bibitem{Ref_PP} N. Troullier and J. L. Martins: Phys. Rev. B\ {\bf 43} (1991) 1993; L. Kleinman and D. M. Bylander: Phys. Rev. Lett.\ {\bf 48} (1982) 1425. 
\bibitem{MaxLoc}N. Marzari and D. Vanderbilt: Phys. Rev. B {\bf 56} (1997) 12847; I. Souza, N. Marzari and D. Vanderbilt: Phys. Rev. B {\bf 65} (2001) 035109.
\bibitem{C4}The broken C4 symmetry for 25\% doping is attributed to the shape of 2x1x1 supercell.
\bibitem{Kontani2} H. Kontani and S. Onari: Phys. Rev. Lett. {\bf 104} (2010) 157001; T. Saito, S. Onari, and H. Kontani: Phys. Rev. B {\bf 82} (2010) 144510.
\bibitem{s+-} I. I. Mazin, D. J. Singh, M. D. Johannes, and M. H. Du: Phys. Rev. Lett. {\bf 101} (2008) 057003; K. Kuroki, S. Onari, R. Arita, H. Usui, H. Kontani, Y. Tanaka and H. Aoki: Phys. Rev. Lett. {\bf 101} (2008) 087004.
\bibitem{Kontani} S. Onari and H. Kontani: Phys. Rev. Lett. {\bf 103} (2009) 177001.
\bibitem{Kariyado} T. Kariyado and M. Ogata: J. Phys. Soc. Jpn. {\bf 79} (2010) 033703.

\end{thebibliography}
\end{document}